%% file: fBs.tex
\documentclass[11pt]{article}
\pdfoutput=1
\usepackage{amsmath,amsfonts,sint,epsfig,cite,graphics}
\usepackage{mathrsfs}
\usepackage{macros_static}
\usepackage{wrapfig}
\usepackage{color}

\begin{document}

\renewcommand{\vec}[1]{{\bf #1}}
\newcommand{\PH}{P_{\rm H}}
\newcommand{\PL}{P_{\rm L}}
\newcommand{\unity}{{I}}

\input titlefBs.tex

\input s1fBs.tex

\input s2fBs.tex

\input s3fBs.tex
\input s4fBs.tex

\vskip2ex\noindent
{\bf Acknowledgements.}
We thank
NIC for allocating computer time
on the APE computers
and the computer farm at DESY, Zeuthen,
and the staff of the computer center at Zeuthen for their support.
This work is supported
by the Deutsche Forschungsgemeinschaft
in the SFB/TR~09,
and by the European community
through EU~Contract No.~MRTN-CT-2006-035482, ``FLAVIAnet''.
N.G. acknowledges financial support from 
the MICINN grant FPA2006-05807,
the Comunidad Aut\'onoma de Madrid programme HEPHACOS~P-ESP-00346,
and participates in the Consolider-Ingenio~2010~CPAN (CSD2007-00042).
T.M. thanks the A.~von~Humboldt~Foundation for support.

\bibliographystyle{JHEP}
\bibliography{spires}

\end{document}

%% file: titlefBs.tex
\begin{titlepage}

\begin{flushright}
\vskip 0.7cm
DESY 09-155\\
SFB/CPP-10-03\\
Edinburgh 2010/15\\
MKPH-T-10-04\\
LPT-Orsay/10-35\\
\end{flushright}

\vskip 0.35cm
\begin{center}
{\Large\bf 
HQET at order $1/m$: III. Decay constants in the quenched approximation}
\end{center}
\vskip 0.35cm
\vbox{
\centerline{
\epsfxsize=2.8 true cm
\epsfbox{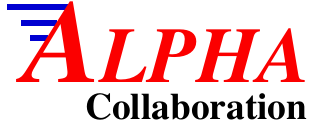}}
}
\vskip 0.1cm
\begin{center}
{
Beno\^it~Blossier$^{\scriptscriptstyle a}$,
Michele~Della~Morte$^{\scriptscriptstyle b}$,
Nicolas~Garron$^{\scriptscriptstyle c,d}$,
Georg~von~Hippel$^{\scriptscriptstyle b,e}$,
Tereza~Mendes$^{\scriptscriptstyle e,f}$, 
Hubert~Simma$^{\scriptscriptstyle e}$,
Rainer~Sommer$^{\scriptscriptstyle e}$ 
}
\vskip 0.5cm
{
\vskip 2.0ex
$^{\scriptstyle a}$
Laboratoire~de~Physique~Th\'eorique,
B\^atiment~210, CNRS et Universit\'e~Paris-Sud~XI,
F-91405~Orsay~Cedex, France
\vskip 2.0ex
$^{\scriptstyle b}$
Institut~f{\"u}r~Kernphysik, University~of~Mainz,
D-55099~Mainz, Germany
\vskip 2.0ex
$^{\scriptstyle c}$
Departamento~de~F\'isica~Te\'orica and Instituto~de~F\'isica~Te\'orica~IFT-UAM/CSIC\\
Universidad~Aut\'onoma~de~Madrid, Cantoblanco 28049~Madrid, Spain
\vskip 2.0ex
$^{\scriptstyle d}$
SUPA, School~of~Physics, University~of~Edinburgh,
Edinburgh~EH9~3JZ, U.K.
\vskip 2.0ex
$^{\scriptstyle e}$
NIC, DESY,
Platanenallee~6, D-15738~Zeuthen, Germany
\vskip 2.0ex
$^{\scriptstyle f}$
IFSC, University~of~S\~ao~Paulo, 
C.P.~369, CEP~13560-970, S\~ao~Carlos~SP, Brazil
}
\vskip 0.775cm
{\bf Abstract}
\vskip 0.1ex
\end{center}
We report on the computation of the $\rm B_{\rm s}$ meson decay constant 
in Heavy Quark Effective Theory on the lattice. The next to leading order
corrections in the HQET expansion are included non-perturbatively. We 
estimate higher order contributions to be very small.
The results are extrapolated to the continuum limit, the main systematic
error affecting the computation is therefore the quenched approximation used 
here. 
The Generalized Eigenvalue Problem and the use of all-to-all propagators
are important technical ingredients of our approach that allow to 
keep statistical and systematic errors under control.
We also report on the decay constant $f_{\rm B'_{\rm s}}$ of the 
first radially excited state in the $\rm B_{\rm s}$ sector, computed in the
static limit.
 
\enlargethispage{2ex}

\end{titlepage}

%% file: s1fBs.tex
\newcommand{\corren}{\varepsilon_{n}}
\newcommand{\corrpn}{\pi_{nn'}}
\newcommand{\eneff}{E_n^\mrm{eff}}
\newcommand{\qeff}{{\cal \hat Q}_n^\mrm{eff}}
\newcommand{\qeffp}{{\cal \hat Q}_{n'}^\mrm{eff}}
\newcommand{\qeffg}{{\cal \hat Q}_1^\mrm{eff}}
\newcommand{\qefff}{{\cal Q}_n^\mrm{eff}}
\newcommand{\qefffp}{{\cal Q}_{n'}^\mrm{eff}}
\newcommand{\qn}{{\cal \hat A}_n}
\newcommand{\qnt}{{\cal \tilde A}_n}
\newcommand{\aeff}{{\cal \hat A}_n^\mrm{eff}}

\section{Introduction}
Flavour physics is becoming a precision field. B-physics measurements may
produce stringent tests of the Standard Model (SM) and consequently 
reveal possible effects coming from New Physics. They are complementary 
to direct searches and they provide constraints on the flavour structure 
of any possible extension of the Standard Model.
At the moment the significance of such tests is limited by the 
uncertainties on the theoretical side~\cite{Antonelli:2009ws}.
A typical example is the process ${\rm B_{\rm s}} \to \mu^+ \mu^-$. 
The SM prediction for the branching ratio is 
O($10^{-9}$)~\cite{BRBsmuplusmuminusSM1,BRBsmuplusmuminusSM2,
BRBsmuplusmuminusSM3} and the best
experimental upper bound (from D0) is $4.2 \times 10^{-8}$ 
@ 90\% CL~\cite{BRBsmuplusmuminusD0}. The decay is very sensitive to an 
extended Higgs sector and may be strongly enhanced in various extensions
of the Standard Model (e.g. the supersymmetric model discussed 
in~\cite{NPBSUSY}). LHCb has a potential to measure a branching ratio
as small as $9\times 10^{-9}$ at 3 $\sigma$
with 0.1 fb$^{-1}$ of data~\cite{BRLHCbBmuplusmuminus}.
The hadronic parameter entering the SM prediction is the $\rm B_{\rm s}$
meson decay constant $f_{\rm B_{\rm s}}$, which is known from the lattice
with an uncertainty of about 
15\%~\cite{DellaMorte:2007ny,Gamizlattice08}.

More precise lattice computations are needed to make progress, however 
heavy quarks on the lattice are difficult due to O$((am_{\rm b})^n)$
discretization errors, where $a$ is the lattice spacing.
A description of heavy-light systems which is suitable for lattice
QCD simulations is given by Heavy Quark Effective Theory 
(HQET)~\cite{Eichten:1989zv,Eichten:1990vp} 
with non-perturbatively determined parameters~\cite{hqet:first1}. 

In this paper we report on a quenched computation of $f_{\rm B_{\rm s}}$ 
performed entirely
in HQET including $1/m_{\rm b}$ corrections non-perturbatively.
The plan of the paper is the following.
In section \ref{secstrategy} we restate the strategy that we have used
 and already explained in
\cite{strategyfBs}, with particular emphasis on the use of the GEVP
variational method~\cite{alphaGEVP}.
In section \ref{secnumerics} we give the numerical values of 
$f_{\rm B_{\rm s}}$ and $f_{\rm B'_{\rm s}}$
obtained at the 3 lattice spacings that we have considered and discuss
the extrapolation  to the continuum limit. We briefly conclude in 
section \ref{secconclusion}.

%% file: s2fBs.tex
\section{Strategy of the computation}\label{secstrategy}

\subsection{Non-perturbative HQET}

We aim at computing  the decay constant $f_{\rm B_{\rm s}}$, 
defined in QCD as
\be
\langle {\rm B}_{\rm s}(\vec{p}=0)|\psibar_{\rm s}\gamma_0\gamma_5 \psi_{\rm b}|0\rangle = f_{\rm B_{\rm s}} m_{\rm B_{\rm s}} \;,
\ee
with the normalization of states
$\langle {\rm B}_{\rm s}(\vec{p})| {\rm B}_{\rm s}(\vec{p}')\rangle=2E(\vec{p}) \delta^3 (\vec{p}-\vec{p}')$,
from matrix elements defined in HQET.
To this end we need to
match  the HQET Lagrangian and the currents to their
QCD counterparts. To order $\minv$, the HQET Lagrangian reads
\bes
\label{e:lhqet}
{\mathscr{L}}_{\rm HQET}(x) 
&=& {\mathscr{L}}_{\rm stat}(x)
    - \omega_{\rm kin} {\cal{O}}_{\rm kin}(x)
    - \omega_{\rm spin} {\cal{O}}_{\rm spin}(x) \,,\\
\label{e:lstat}
{\mathscr{L}}_{\rm stat}(x) &=&
\heavyb(x) \left(D_0 + \dmstat \right)\psi_{\rm h}(x)\,, \\
\label{e:ofirst}
{\cal{O}}_{\rm kin}(x) &=& \heavyb(x) \vecD^2\psi_{\rm h}(x)\,,\quad
{\cal{O}}_{\rm spin}(x) \;=\; \heavyb(x) \vecsigma \cdot \vecB \psi_{\rm h}(x)\,,
\ees
where $\psi_{\rm h}$ satisfies $\frac{1+\gamma_0}{2}\psi_{\rm h}=\psi_{\rm h}$,
and $\omega_{\rm kin}$ and $\omega_{\rm spin}$ are matching parameters
whose tree-level values are $\omega_{\rm kin}=\omega_{\rm spin}=1/(2m_{\rm b})$,
and $\dmstat$ is a counter-term that absorbs the power-divergences of
the static quark self energy.

Again to order $\minv$, the time-component of the QCD axial current $A_0^{\rm QCD}(x) = \psibar_{\rm s}(x)\gamma_0\gamma_5 \psi_{\rm b}(x)$ 
corresponds to the effective current
\bes
A_0^{\rm HQET}(x) &=& Z^{\rm HQET}_{\rm A} [A_0^{\rm stat}(x) + 
\sum_{i=1}^2 c_{\rm A}^{(i)}A_0^{(i)}(x)] \;, \\
A_0^{(1)}(x) &=& \psibar_{\rm s} \frac{1}{2}\gamma_5\gamma_i(\nabla_i^{\rm S}
-\overleftarrow{\nabla}_i^{\rm S})\psi_{\rm h}(x) \; , \label{e:impcur}\\
A_0^{(2)}(x) &=& -\tilde{\partial}_iA_i^{\rm stat}(x) \;, \quad
A_i^{\rm stat}(x) = \psibar_{\rm s}(x) \gamma_i\gamma_5 \psi_{\rm h}(x) \, ,
\ees
where all derivatives are symmetrized
\be
\tilde{\partial}_i=\frac{1}{2}(\partial_i+\partial_i^*)\;, \quad
\nabla^{\rm S}_i=\frac{1}{2}(\nabla_i+\nabla_i^*)\;, \quad
\overleftarrow{\nabla}^{\rm S}_i=\frac{1}{2}(
\overleftarrow{\nabla}_i+\overleftarrow{\nabla}_i^*) \;.
\ee
The renormalization constant $Z_{\rm A}^{\rm HQET}$ depends on 
the ratio $m_{\rm s}/m_{\rm b}$. This is a small effect, which is 
further reduced by a factor of the coupling constant $\alpha(m_{\rm b})$.
We will ignore this dependence and use the value of $Z_{\rm A}^{\rm HQET}$
determined with a  massless light quark~\cite{hqet:first1}.
Note in addition that the operator $A_0^{(2)}$ does not 
contribute to correlation functions and matrix elements at zero spatial
momentum, such as those we are interested in here.

At the static order the Lagrangian is automatically $\rmO(a)$ improved, 
therefore the
current and its on-shell matrix elements are $\rmO(a)$ improved if one 
sets $c_{\rm A}^{(1)}=a c_{\rm A}^{\rm stat}$, 
where $c_{\rm A}^{\rm stat}$ is the
improvement coefficient of the static-light axial current introduced 
in~\cite{Kurth:2000ki}. % For the discretizations of the static action used
% here the value of $c_{\rm A}^{\rm stat}$ is known to O($g_0^2)$ 
% from~\cite{Cstatpalombi}.
%
When $\rmO(\minv)$ corrections are included,
the only terms linear in $a$ that are introduced are accompanied by
a factor $\minv$, so that the leading discretization errors are
$\rmO(a/m_{\rm b},\, a^2)$.

In order to retain the renormalizability of the static theory
also at $\rmO(\minv)$, we treat the theory in a strict expansion
in $\minv$, where the $\rmO(\minv)$ parts of the action are inserted
in correlations functions that are computed in the static
approximation. 
As new divergences appear at each order in the expansion, 
the renormalization
constants are also expanded in $\minv$, i.e. 
$\log Z_{\rm A}^{\rm HQET}=\log Z_{\rm A}^{\rm stat}+ \log Z_{\rm A}^{1/m}$, 
and all terms quadratic in 
$\minv$ are consistently dropped. 

As long as we restrict our studies to the decay constants only, to fully
 specify HQET  the parameters  $\delta m$, $\omega_{\rm kin}$,
$\omega_{\rm spin}$, $Z^{\rm HQET}_{\rm A}$, and $c_{\rm A}^{(1)}$ must
be determined by matching the effective theory to QCD.
Using the Schr\"odinger functional, our collaboration has performed a
fully non-perturbative determination of the parameters of HQET
\cite{hqet:first1}.
Here we employ the same discretization of QCD and HQET
and in particular
use the determined values for the parameters of the effective theory.

\subsection{The Generalized Eigenvalue Problem}
\def\first{{1/m}}
\def\stat{\mrm{stat}}

We follow here the application of the GEVP~\cite{gevp:michael,phaseshifts:LW}
described in~\cite{alphaGEVP}.
For the sake of completeness we recall the basic ingredients of the
 method.
The matrix of Euclidean space correlation functions
between the zero-momentum projection $a^3\sum_\vecx O_i(x)=\tilde O_i(x_0)$
of some local composite fields $O_i(x)$ with the spectral representation
\bes
  \label{e:cij}
  C_{ij}(t) &=& \langle \tilde O_i(t) \tilde O_j^*(0) \rangle =
  \sum_{n=1}^\infty \rme^{-E_n t} \psi_{ni}\psi_{nj}^*\,,\quad
  i,j = 1,\ldots,N  \\ && \quad \psi_{ni} \equiv (\psi_n)_i =
  \langle 0|\hat O_i|n\rangle
  \,,\quad
  E_n < E_{n+1} \,,
 \nonumber
\ees
provides the basis for the Generalized Eigenvalue Problem (GEVP)
\bes \label{e:gevp}
  C(t)\, v_n(t,t_0) = \lambda_n(t,t_0)\, C(t_0)\,v_n(t,t_0) \,,
  \quad n=1,\ldots,N\,,\quad t>t_0\,.
\ees
An effective creation operator for the $n^{{\rm th}}$ state
can be defined by
\bes
    \left.\qeff\right.^\dagger(t,t_0) &=& R_n(t,t_0) \,(v_n(t,t_0)\,,\,\hat{O}^\dagger) \,,\\
    \label{Rn}
    R_n(t,t_0) &=&
               \left(v_n(t,t_0)\,,\, C(t)\,v_n(t,t_0)\right)^{-1/2}
               \left(\lambda_n(t_0+a,t_0) \over \lambda_n(t_0+2a,t_0)\right)^{t/(2a)}\,, 
%%%    (u,w) &=& \sum_{i=1}^N u^*_i w_i \,.
\ees
with
\be
    (u,w) = \sum_{i=1}^N u^*_i w_i \,.
\ee

Defining the vector of correlators of a composite field $P$  
(which does not have to be among the $\tilde O_i$)
\be
C_{{\rm P},i}(t) = \langle P(t) \tilde O_i^*(0)\rangle \,,\quad i = 1,\ldots,N \,,
\ee
the effective matrix elements
\be
   p_n^\mrm{eff}(t,t_0) = R_n(t,t_0)\, (v_n(t,t_0)\,,\,C_{\rm{P}}(t)\,)\,,
\ee
approximate the matrix elements of the corresponding operator $\hat{P}$ as
\bes
    p_n^\mrm{eff}(t,t_0) & = & \ketbra{0}{\hat{P}}{n} + \pi(t,t_0)\,,\\
    \pi(t,t_0) & = & \rmO\left(\rme^{-(E_{N+1}-E_n)t_0}\right)\,.
    \label{perror}
\ees
The definition of $R_n$ in eq.~(\ref{Rn}) is slightly different from the one in
\cite{alphaGEVP}
and has the advantage of being defined at all
(and not only even) values of $t$, thus giving better statistical precision
for the final result while preserving the same control (\ref{perror}) over the contamination
from excited states as proven in \cite{alphaGEVP}.

After expanding the correlators to first order in 
$\omega\sim\minv$
\bes
   C(t)         & = & C^\stat(t) +\omega\, C^\first(t) +\rmO(\omega^2)\,,\\
   C_{\rm{P}}(t) & = & C_{\rm{P}}^\stat(t) + \omega\, C_{\rm{P}}^\first(t) + \rmO(\omega^2)\,,
\ees
we consider the GEVP in perturbation theory in $\minv$ and find
\bes
p_n^\mrm{eff}(t,t_0) &=& p_n^\mrm{eff,stat}(t,t_0)\left(1+\omega\, p_n^{{\rm eff},1/m}(t,t_0)+\rmO(\omega^2)\right)\,, \\
p_n^{{\rm eff},1/m}(t,t_0) &=& {R_n^{1/m} \over R_n^\mrm{stat}}\,+\,
   {(v_n^\stat, C_\mrm{P}^\first(t)) \over (v_n^\stat,C_\mrm{P}^\stat(t))} \,+\,
   {(v_n^\first,C_\mrm{P}^\stat(t)) \over (v_n^\stat,C_\mrm{P}^\stat(t))}\,, \nonumber
\ees
where
\be
{R^{1/m}_n \over R^\mrm{stat}_n} = -{1 \over 2}
{(v^\mrm{stat}_n, C^{1/m}(t) v^\mrm{stat}_n) \over
(v^\mrm{stat}_n, C^\mrm{stat}(t) v^\mrm{stat}_n)} + {t \over 2a}
\left({\lambda^{1/m}_n(t_0+a,t_0)\over\lambda^\mrm{stat}_n(t_0+a,t_0)}
- {\lambda^{1/m}_n(t_0+2a,t_0)\over\lambda^\mrm{stat}_n(t_0+2a,t_0)}\right)\,,\nonumber
\ee\bes
{\lambda_n^\first(t,t_0)\over \lambda_n^\stat(t,t_0)}
  &=& \left(v_n^\stat\,,\,
             [[\lambda_n^\stat(t,t_0)]^{-1}\,C^{\first}(t)- C^{\first}(t_0)]
             v_n^\stat\right),\\
v_n^{1/m} &=& \sum_{k=1, k\ne n}^N v_k^\stat
      { \left({v}_k^\stat, [C^{1/m}(t)-\lambda_n^\stat(t,t_0)C^{1/m}(t_0)]\,
         {v}_n^\stat\right) \over
               \lambda_n^\stat(t,t_0)-\lambda_k^\stat(t,t_0) }\,.\nonumber
\ees
Thus, in order to obtain the effective matrix elements, the GEVP has to 
be solved for the static correlation functions only
\bes
C^\stat(t)\,v_n^\stat=\lambda_n^\stat(t,t_0)\,C^\stat(t_0)\,v_n^\stat\,,
\quad  v_n^\stat\equiv v_n^\stat(t,t_0)\;.
\ees

With these definitions, and by organizing the $\minv$ expansion in the way we 
discussed in the previous section, the decay constant of a pseudoscalar $\rm B_s$ meson ($n=1$) 
or of radial excitations ($n>1$) computed in the static approximation and in HQET 
(i.e. including terms of order $1/m_{\rm b}$), respectively, read 
\bea\label{eqfbs}
f^{\rm stat}_n \sqrt{m_n/2}
&=&Z_{\rm A}^{\rm stat} 
\,(1+b_{\rm A}^{\rm stat}am_{\rm q})\,p^{\rm stat}_n \left( 1 + c_{\rm A}^{\rm stat} p^{{\rm A}^{(1)}}_n \right)\; ,\\
f^{\rm HQET}_n \sqrt{m_n/2}
&=&Z_{\rm A}^{\rm HQET} 
\,(1+b_{\rm A}^{\rm stat}am_{\rm q})\,p^{\rm stat}_n \left(1+\omega_{\rm kin}\, p^{\rm kin}_n +
\omega_{\rm spin}\, p^{\rm spin}_n + c_{\rm A}^{(1)} 
p^{{\rm A}^{(1)}}_n\right), \nonumber
\eea
where $p^{\rm stat}_n$, $p^{\rm kin}_{n}$, $p^{\rm spin}_n$ and
$p^{{\rm A}^{(1)}}_n$ are the plateau values of the corresponding
effective matrix elements (see~\cite{alphaGEVP} where however 
$p^{{\rm A}^{(1)}}_n$ is called $p^{\delta {\rm A}}_n$).
For the improvement term proportional to $b_{\rm A}^{\rm stat}am_{\rm q}$
we use the 1-loop estimates of the coefficient 
$b_{\rm A}^{\rm stat}$ from~\cite{Cstatpalombi}. In the formulae 
$am_{\rm q}$ is the bare subtracted strange quark mass 
$\frac{1}{2}\left(\frac{1}{\kappa_{\rm s}}-
\frac{1}{\kappa_{\rm c}}\right)$, with
$\kappa_{\rm c}$ the critical value of the hopping parameter defined
through the vanishing of the quark mass derived from the axial 
Ward identity.

In order to consistently truncate the expansion at order $\minv$,
it is convenient to take the logarithm of (\ref{eqfbs}) and expand the
logarithms (rather than expanding directly the product of the factors
from the correlation function times its renormalization constant).

%% file: s3fBs.tex
\section{Numerical results}\label{secnumerics}

\subsection{Simulation parameters}

We are now ready to present the result of our numerical
simulations to extract $f_{\rm B_s}$. 
The parameters of the
simulations are given in Table \ref{tabparamsimul}.
Each ensemble contains 100 quenched configurations.
The heavy quark is described by the HYP1 and HYP2 static actions
\cite{HYP,HYP:pot,Dellam2005}
while the valence strange quark is described by the non-perturbatively 
${\cal O}(a)$-improved Wilson action
\cite{impr:SW, impr:pap3}.
Our lattices are $L^3\times T$ with $L\approx 1.5$ fm, $T=2L$, and
periodic boundary conditions are applied in all directions.
We use all-to-all propagators based on the Dublin method
\cite{dublin},
but with even-odd preconditioning and $N_{\rm L}$ approximate
(instead of exact) low modes; for details of our method the
reader is referred to
\cite{hqet:first3}.
No low modes have been computed for $\beta=6.4956$ because
the numerical cost would have been too high with respect to the gain
in statistical precision; instead, we have improved the statistics by
using $N_\eta=4$ stochastic noises, twice the number of noise sources
used at the other lattice spacings.
\begin{table}
\begin{center}
\begin{tabular}{lrlllrr}
\hline\hline
$\beta$&$r_0/a$&$L^3 \times T$&$\kappa_{\rm s}$&$\kappa_{\rm c}$&
$N_{\rm L}$&$N_\eta$\\
\hline
6.0219&5.57&$16^3\times 32$&0.133849&0.135081&50&2\\
6.2885&8.38&$24^3\times 48$&0.1349798&0.135750&50&2\\
6.4956&11.03&$32^3\times 64$&0.1350299&0.135593&0&4\\
\hline\hline
\end{tabular}
\end{center}
\caption{\label{tabparamsimul} Parameters of the simulations:
inverse coupling $\beta$,
approximate scale parameter $r_0$ in lattice units
\protect\cite{r0formula},
spacetime volume,
hopping parameter corresponding to the strange quark mass
\protect\cite{mbar:pap3},
critical hopping parameter
\protect\cite{Dellam2005},
and numbers of low-lying eigenmodes and stochastic noises used.}
\end{table}

\subsection{Bare matrix elements}

In Table \ref{tabnumhadmatelmts} we give the numerical values of the
bare hadronic matrix elements entering
the formulae in eq. (\ref{eqfbs}) for $f_{\rm B_s}\equiv f^{\rm HQET}_1$
and $f^{\rm stat}_{\rm B_s'}\equiv f^{\rm stat}_2$
at each of our three lattice spacings
for both the HYP1 and the HYP2 static quark action.

\begin{table}
\begin{center}
\begin{tabular}{llllll}
\hline\hline
        &            &\multicolumn{2}{c}{HYP1}&\multicolumn{2}{c}{HYP2} \\
$\beta$ & Observable             & Fit       & Plateau  & Fit      & Plateau \\
\hline
6.0219  &$a^{3/2}p^{\rm stat}_1$ & 0.1424(5) & 0.1429(9)& 0.1238(4)& 0.1242(8)\\
        &$a^{3/2}p^{\rm stat}_2$ & 0.204(5)  & 0.203(4) & 0.164(4) & 0.164(3)\\
        &$ap^{\rm kin}_1$        & -1.46(1)  & -1.46(1) & -0.802(9)& -0.802(8)\\
        &$ap^{\rm spin}_1$       & 0.421(2)  & 0.423(6) & 0.408(2) & 0.409(5)\\
        &$ap^{\rm A^{(1)}}_1$   & 0.4186(6) & 0.420(1) & 0.3755(5)& 0.376(1)\\
        &$ap^{\rm A^{(1)}}_2$   & 0.615(4)  & 0.614(4) & 0.599(5) & 0.599(5)\\[0.5ex]\hline
6.2885  &$a^{3/2}p^{\rm stat}_1$ & 0.0767(2) & 0.0771(7)& 0.0690(2)& 0.0692(6)\\
        &$a^{3/2}p^{\rm stat}_2$ & 0.099(1)  & 0.102(5) & 0.085(1) & 0.086(4)\\
        &$ap^{\rm kin}_1$        & -1.069(6) & -1.07(1)& -0.604(5)& -0.61(1)\\
        &$ap^{\rm spin}_1$       & 0.401(2)  & 0.401(3) & 0.386(1) & 0.386(2)\\
        &$ap^{\rm A^{(1)}}_1$   & 0.3524(3) & 0.3532(9) & 0.3122(3)& 0.313(3)\\
        &$ap^{\rm A^{(1)}}_2$   & 0.494(2)  & 0.492(7) & 0.460(2) & 0.458(3)\\[0.5ex]\hline
6.4956  &$a^{3/2}p^{\rm stat}_1$ & 0.0491(2) & 0.0499(5)& 0.0448(2)& 0.0455(4)\\
        &$a^{3/2}p^{\rm stat}_2$ & 0.0659(9) & 0.066(3) & 0.059(1)& 0.059(3)\\
        &$ap^{\rm kin}_1$        & -0.97(1)  & -0.97(3) & -0.51(1)& -0.48(3)\\
        &$ap^{\rm spin}_1$       & 0.365(3)  & 0.368(8) & 0.353(3) & 0.354(6)\\
        &$ap^{\rm A^{(1)}}_1$   & 0.3095(5) & 0.311(1) & 0.2719(4)& 0.273(1)\\
        &$ap^{\rm A^{(1)}}_2$   & 0.424(2)  & 0.423(6) & 0.386(2) & 0.386(8)\\[0.5ex]
\hline\hline
\end{tabular}
\end{center}
\caption{\label{tabnumhadmatelmts} Bare matrix elements involved in the
decay constants of the $\rm B_s$ ground state
(in HQET to order $\minv$) and first radial excitation (at static order).}
\end{table}

The interpolating fields are constructed using quark bilinears
\begin{eqnarray}
O_k(x) &=& \psibar_{\rm h}(x)\gamma_0\gamma_5\psi_{\rm l}^{(k)}(x) \,, \\
O_k^*(x) &=& \psibar_{\rm l}^{(k)}(x)\gamma_0\gamma_5\psi_{\rm h}(x) \,,\nonumber
\end{eqnarray}
built from the static quark field $\psi_{\rm h}(x)$ and different levels
of Gaussian smearing
\cite{wavef:wupp1}
for the light quark field with APE smeared links
\cite{smear:ape,Basak:2005gi}
in the Laplacian
\begin{equation}
\psi_{\rm l}^{(k)}(x) = \left( 1+\kappa_{\rm G}\,a^2\,\Delta \right)^{R_k} \psi_{\rm l}(x) \,,
\end{equation}
with exactly the same parameters as in \cite{hqet:first3}.

For these bilinears, we compute the following correlators:
\bea
\nonumber
C^{\rm{stat}}_{ij}(t)& = &
\sum_{x,\bf{y}}
\left< O_i(x_0+t,\vecy)\,O^*_j(x)\right>_\stat,\nonumber\\
C^{\rm{kin/spin}}_{ij}(t)& = &
\sum_{x,\vec{y},z}
\left< O_i(x_0+t,\vecy)\,O^*_j(x)\,  O_{\rm kin/spin}(z)\right>_\stat \,,
\\
\nonumber
C^{\rm{stat}}_{A^{(1)},i}(t)& = &
\sum_{x,\vec{y}}
\left< A^{(1)}_0(x_0+t,\vec{y}) O^*_i(x)\right>_\stat \,,
\eea
where the $\rmO(\minv)$ fields and $A^{(1)}_0$ have been
defined in \eq{e:ofirst} and \eq{e:impcur}.

\begin{figure}
\includegraphics[width=0.5\textwidth,keepaspectratio=]{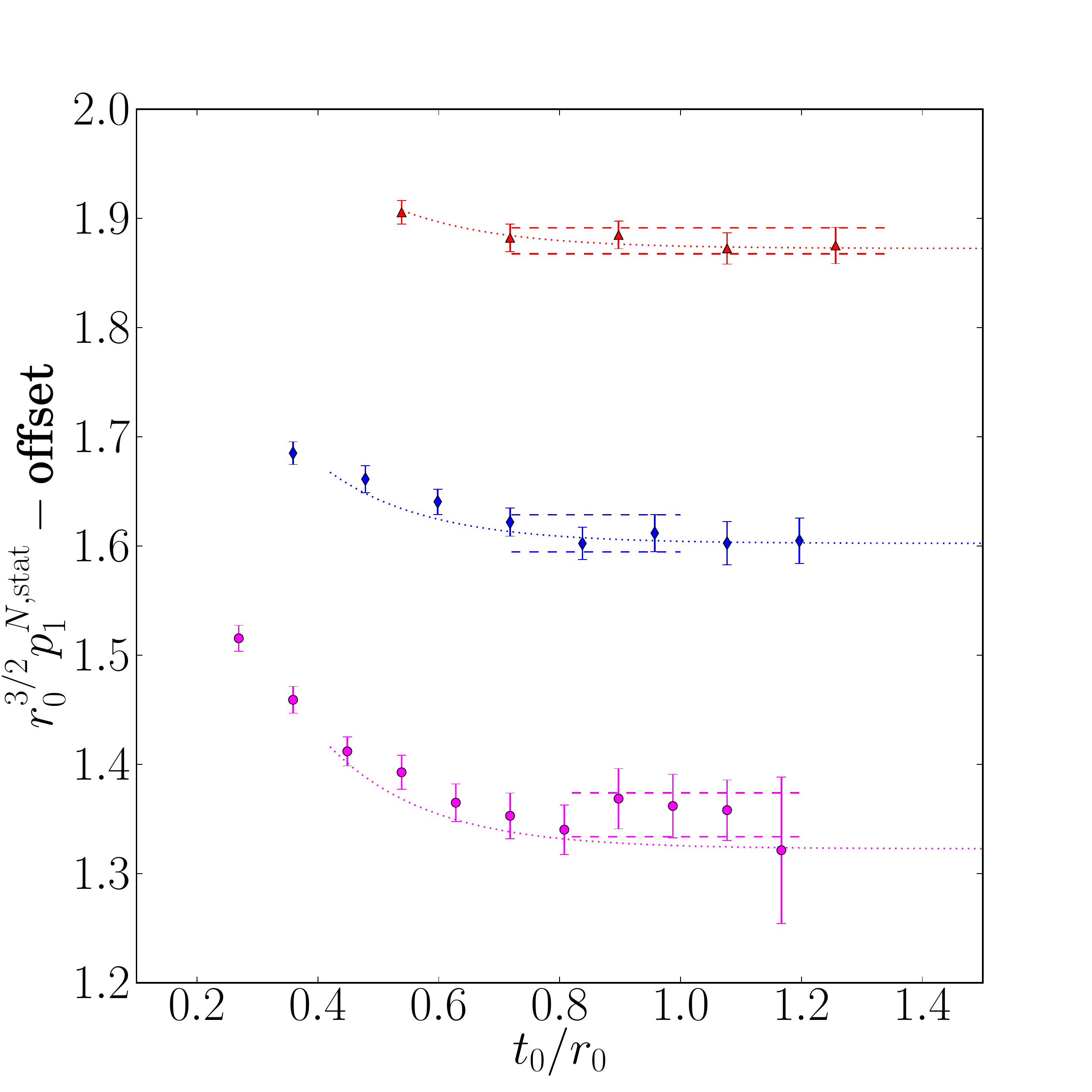}
\includegraphics[width=0.5\textwidth,keepaspectratio=]{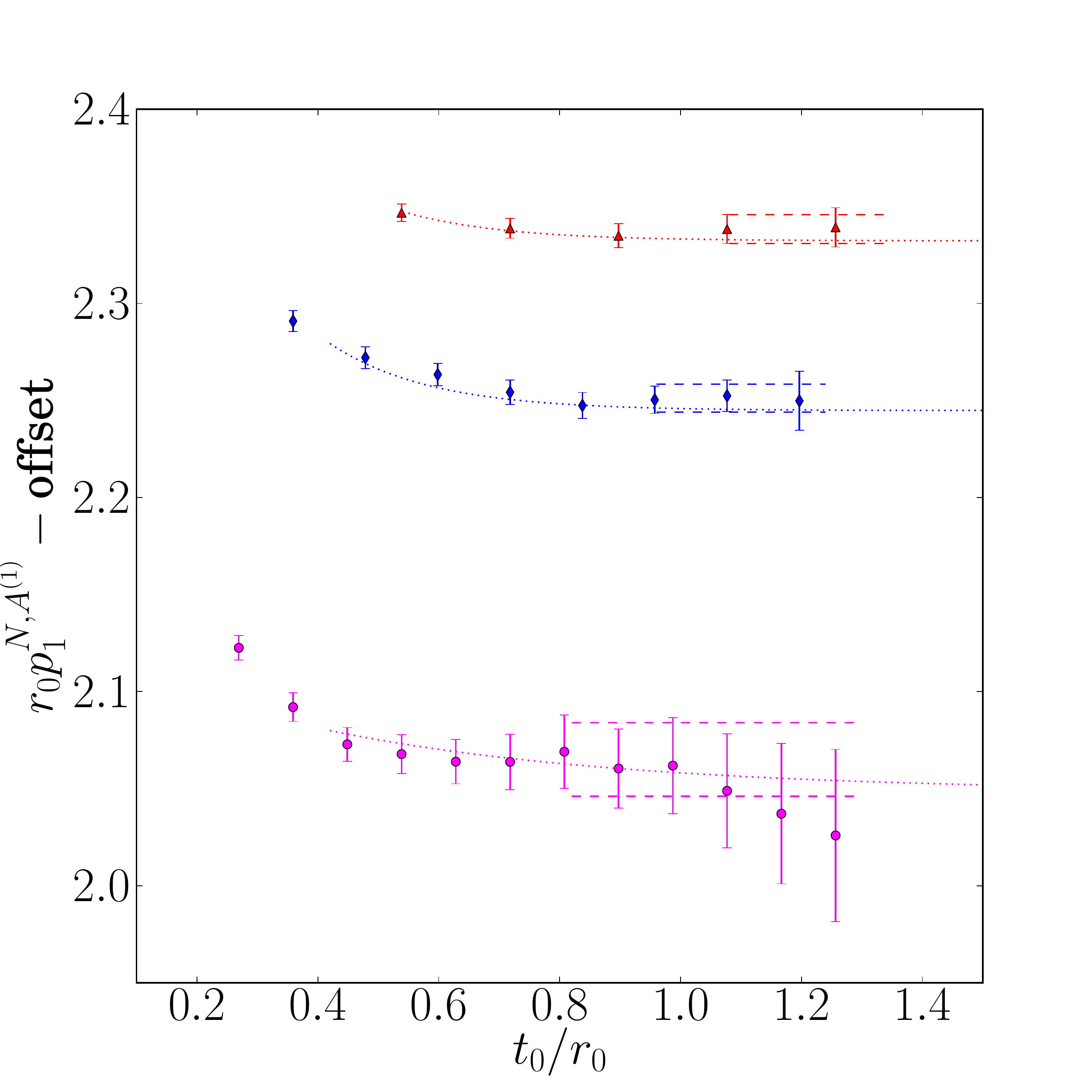}\\
\includegraphics[width=0.5\textwidth,keepaspectratio=]{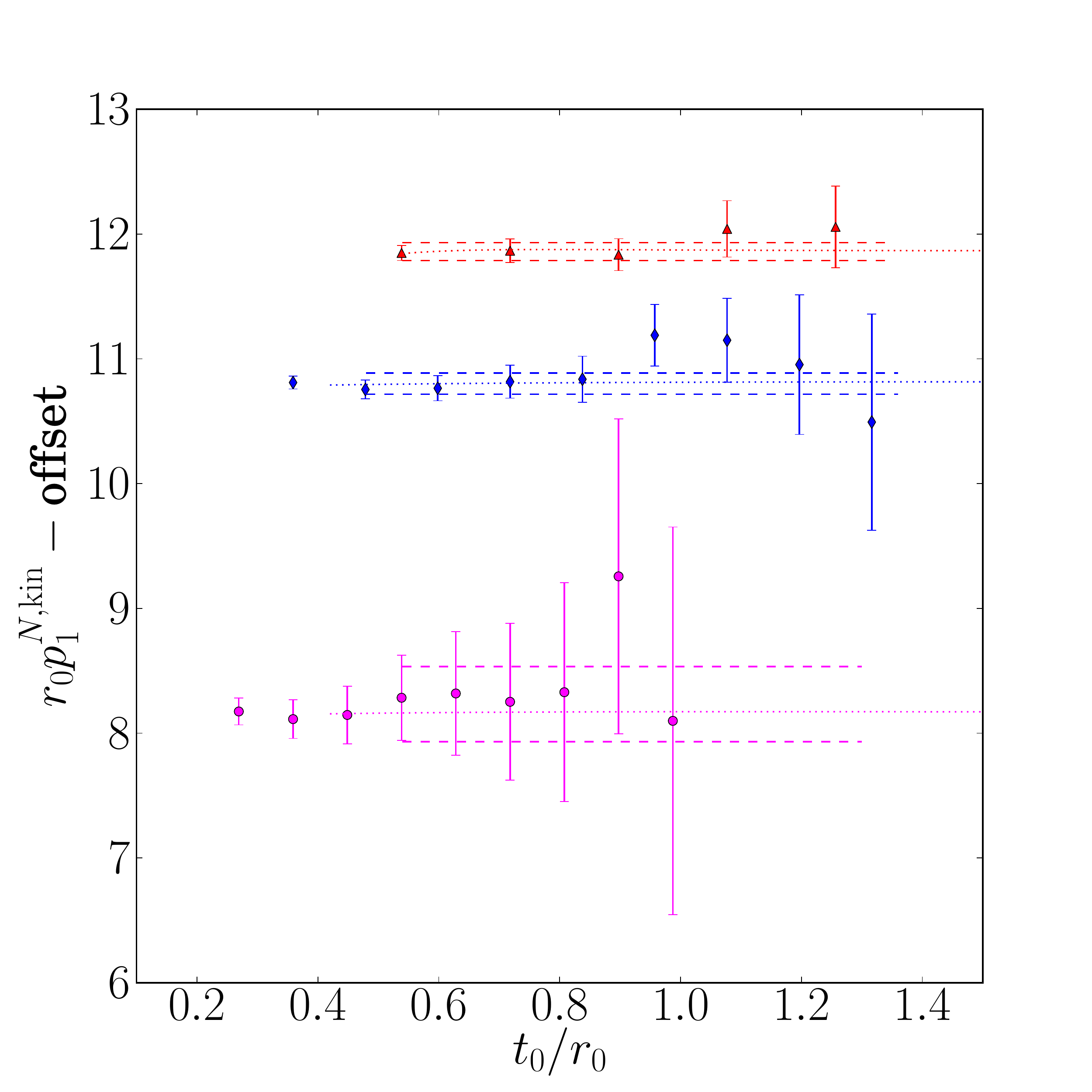}
\includegraphics[width=0.5\textwidth,keepaspectratio=]{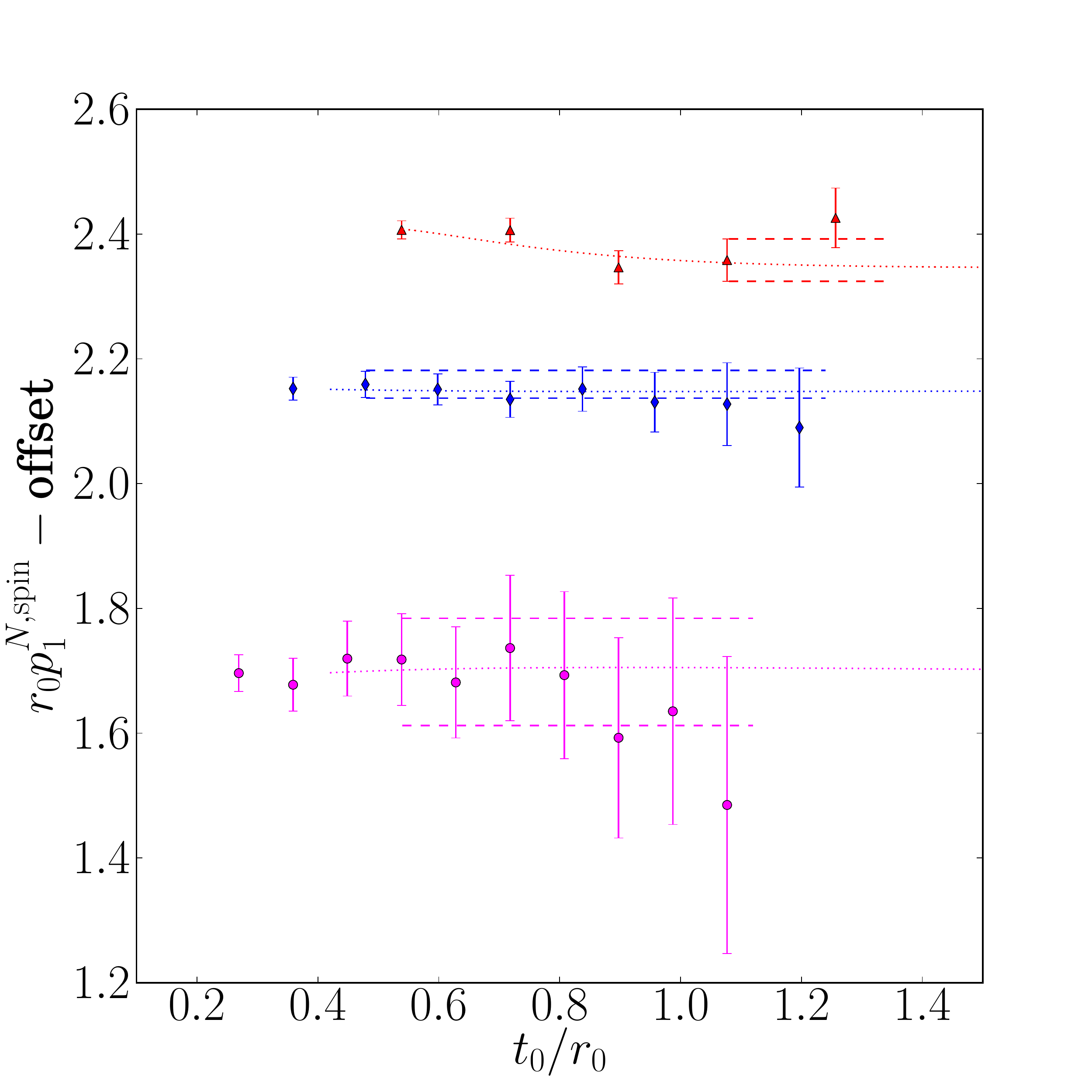}
\caption{Illustration of some plateaux. Top:
$p_{1}^{N, \rm stat}$ (left) and $p_{1}^{N, A^{(1)}}$ (right);
bottom $p_{1}^{N, \rm kin}$ (left) and $p_{1}^{N, \rm spin}$ (right).
In each plot, the lattice spacing is decreasing
from top to bottom. Dotted lines represent the global fit,
while dashed lines indicate the chosen plateau.
In the plots, $N=5$ and $t-t_0=3a$.}
\label{fig:plateaux}
\end{figure}

We have followed the procedure explained in detail in
\cite{hqet:first3}
to choose the time ranges over which
we fit the various plateaux.
Some examples of the plateaux found are shown in
Figure \ref{fig:plateaux}; it can be seen that
without some knowledge of the analytical form
of the leading corrections it would often be difficult
to tell whether a reliable plateau has been found.

We first fit the matrix elements to the expected form
\bea
p_n^{N,\rm stat}(t,t_0) &=& p_n^{\rm stat}
+ \gamma_{n,N}^{\rm stat}\,\rme^{-(E^{\rm stat}_{N+1}-E^{\rm stat}_n)t_0} \,, \nonumber \\
p_{n}^{N,\rm x}(t,t_0) &=& p_n^{\rm x} +\left[
\gamma_{n,N}^{\rm x}
-{\gamma_{n,N}^{\rm stat}\over p_n^{\rm stat}}
\,t_0\,(E^{\rm x}_{N+1}-E^{\rm x}_n)
\right]\rme^{-(E^{\rm stat}_{N+1}-E^{\rm stat}_n)t_0} \,, \nonumber \\
p_n^{N,A^{(1)}}(t,t_0) &=& p_n^{A^{(1)}} +
\gamma_{n,N}^{A^{(1)}} \,\rme^{-(E^{\rm stat}_{N+1}-E^{\rm stat}_n)t_0} \,,
\eea
(where ${\rm x}\in\{{\rm kin},{\rm spin}\} $)
using the energy levels extracted by the procedure described
in
\cite{hqet:first3}
as input parameters. Then, in a second step, we form plateau averages
starting from $t_0=t_{0,\rm min}$ at each value of $N$ and $\Delta t=t-t_0$,
and take as our final estimate that plateau for which the
sum $\sigma_{\rm tot}=\sigma_{\rm stat}+\sigma_{\rm sys}$
of the statistical error $\sigma_{\rm stat}$ of the plateau average and
the maximum systematic error $\sigma_{\rm sys}=\pi(t,t_{0,\rm min})$
becomes minimal,
subject to the constraint that $\sigma_{\rm sys}<\frac13 \sigma_{\rm stat}$.
We impose the latter constraint in order to ensure that the
total error is dominated by statistical errors.
The extracted bare matrix elements are given in table
\ref{tabnumhadmatelmts}, quoting not only the final plateau average,
but also the result of the fit, which generally agrees rather well with
the final result.

\subsection{Continuum limit}

From the bare matrix elements and the parameters of HQET,
determined in 
\cite{hqet:first1}
with the HYP1 and HYP2 static actions, we form dimensionless quantities 
\bes\label{eqn:phis}
\tilde\Phi^\stat_n =& \log\left(r_0^{3/2}\Phi^\stat_n\right/\sqrt{2}) &= \log Z_{\rm A}^\stat + \log(r_0^{3/2} p^\stat_n) \;,\\
\tilde\Phi^{\stat,\rm imp}_n =& \log\left(r_0^{3/2}\Phi^{\stat,\rm imp}_n/\sqrt{2}\right) &= \log Z_{\rm A}^{\stat,\rm imp} + \log(r_0^{3/2} p^\stat_n) \\ &&+ c_{\rm A}^{\rm stat} a p_n^{A^{(1)}} +
b_{\rm A}^{\rm stat} am_{\rm q} \;,\nonumber\\
\tilde\Phi^{\rm HQET}_1 =& \log\left(r_0^{3/2}\Phi^{\rm HQET}_1/\sqrt{2}\right) &= \tilde\Phi^\stat_1 + b_{\rm A}^{\rm stat} am_{\rm q} + \log Z_{\rm A}^\first \\ &&+ \omega_{\rm k in}p_1^{\rm kin} + \omega_{\rm spin}p_1^{\rm spin} + c_{\rm A}^{(1)} p_1^{A^{(1)}} \,,\nonumber
\ees
in which the divergences cancel exactly to order $\rmO(\minv)$, and use them to
compute
%%% $\Phi_n=f^{(n)}_{\rm B_s}\sqrt{m^{(n)}_{\rm B_s}}$
$f^{\rm x}_{\rm B_s} = \Phi^{\rm x}_1/\sqrt{m_{\rm B_s}}$.

For a comparison to the static limit and previous work, we also consider
\bes
\tilde\Phi^{\rm RGI}_1 =& \log\left(r_0^{3/2}\Phi^{\rm RGI}_1/\sqrt{2}\right) &= \log Z_{\rm A,RGI}^{\rm stat} + \log(r_0^{3/2} p^\stat_1) + c_{\rm A}^{\rm stat} a p_1^{{\rm A}^{(1)}} \;,
\ees
where $Z_{\rm A,RGI}^{\rm stat}$ is the renormalization factor of
the Renormalization Group Invariant static-light axial current,
as defined in~\cite{KurthHeitger2003}. 
In contrast to $Z_{\rm A,RGI}^{\rm stat}$, the HQET parameter
$Z_{\rm A}^{\rm stat}$ in eq. (\ref{eqn:phis}) has been determined by a
non-perturbative matching at finite mass.
The correspondence is
\begin{equation}
Z_{\rm A}^{\rm stat}=Z_{\rm A,RGI}^{\rm stat}\,C_{\rm PS}(M_{\rm b}/\Lambda)\; ,
\end{equation}
in terms of the conversion function $C_{\rm PS}$ introduced
in~\cite{KurthHeitger2003} and now known up to three-loops
\cite{HeitgerWennekers2004,Chetyrkin:2003vi,Bekavac:2009zc}.
For $Z_{\rm A,RGI}^{\rm stat}$ we use the non-perturbative
value from~\cite{Dellam2005}.

Since both of the static actions used are discretizations
of the same continuum theory, we perform a combined continuum
limit by fitting a function of the form ($k=1,2$ for HYP1, HYP2 actions)
\be
\Psi_{i,k}(a/r_0) = A_i + B_{i,k}\cdot (a/r_0)^{s_i} \,.
\ee
For $\Psi_i\in\{\tilde\Phi^{\stat,\rm imp}_1,
r_0^{3/2}\Phi^{\stat,\rm imp}_1, \tilde\Phi^{\rm RGI}_1,
r_0^{3/2}\Phi^{\rm RGI}_1\}$, we use $s_i=2$ because the static axial 
current has been $\rmO(a)$-improved using
the coefficients $c_{\rm A}^{\rm stat}$ given in
\cite{Cstatpalombi}
for the actions HYP1 and HYP2. For $\Psi_i\in\{\tilde\Phi^{\rm HQET}_1,
r_0^{3/2}\Phi^{\rm HQET}_1\}$, the $\rmO(a)$ corrections are suppressed
by $\minv$, and given the flatness of the observables in $a$, we feel
justified in employing $s_i=2$ in this case, too.

To estimate the systematic error on $r_0^{3/2}\Phi^{\stat,\rm imp}_1$
incurred from using the one-loop value
of $c_{\rm A}^{\rm stat}$, we compute the continuum limit also for
$c_{\rm A}^{\rm stat}=0$ using a quadratic extrapolation
and compare the result to the continuum limit obtained using
the one-loop value. As can be seen from table \ref{tab:mat_elems},
the influence of $c_{\rm A}^{\rm stat}$ is negligible at this level.

Statistical errors are computed by a jackknife analysis that also
includes the correlation among HQET parameters.
We find that the results obtained from taking the continuum limit of
$\tilde\Phi^{\rm x}_n$ and using it to compute $\Phi^{\rm x}_n$, and
from taking the continuum limit of $\Phi^{\rm x}_n$ directly agree
well within the errors.

\begin{table}
\begin{center}
\begin{tabular}{llrrr|l}
\hline\hline
  & &  $\beta=6.0219$ & $\beta=6.2885$ & $\beta=6.4956$ & cont. limit\\\hline
$r_0^{3/2}\Phi^\stat_1$ & HYP1 & 2.30(4) & 2.22(4) & 2.19(4) & 2.14(4) \\
                        & HYP2 & 2.19(3) & 2.16(4) & 2.15(4) & \\\hline
$r_0^{3/2}\Phi^{\stat,\rm imp}$& HYP1 & 2.31(3) & 2.22(3) & 2.19(4) & 2.15(4) \\
                               & HYP2 & 2.23(3) & 2.18(3) & 2.16(3) & \\\hline
$r_0^{3/2}\Phi^{\rm HQET}_1$ & HYP1 & 1.96(4) & 2.02(4) & 1.94(5) & 2.02(4) \\
                             & HYP2 & 2.00(3) & 2.02(4) & 2.04(5) & \\\hline
$r_0^{3/2}\Phi^{\rm RGI}_1$ & HYP1 & 1.95(3) & 1.87(3) & 1.85(3) & 1.80(3) \\
($c_{\rm A}^{\rm stat}=0$)  & HYP2 & 1.87(3) & 1.83(3) & 1.81(3) & \\\hline
$r_0^{3/2}\Phi^{\rm RGI}_1$    & HYP1 & 1.96(3) & 1.88(3) & 1.85(3) & 1.81(3) \\
(1-loop $c_{\rm A}^{\rm stat}$)& HYP2 & 1.90(3) & 1.84(3) & 1.82(3) &\\
\hline\hline
\end{tabular}
\end{center}
\caption{The matrix elements in units of the scale $r_0$.
Shown are the results at each $\beta$ for both static-quark actions,
together with their common continuum limit.
Note that to compare $\Phi_1^{{\rm RGI}}$ with the first
three continuum-limit results one has to consider the combination
$C_{\rm PS} \times \Phi_1^{{\rm RGI}} = 2.20(4)$
(1-loop $c_{\rm A}^{\rm stat}$).
Here the conversion function $C_{\rm PS}$
\protect\cite{KurthHeitger2003, HeitgerWennekers2004}
has been computed with the three-loop anomalous dimension
from~\protect\cite{Chetyrkin:2003vi,Bekavac:2009zc}.
\label{tab:mat_elems}}
\end{table}
\begin{figure}
\begin{center}
\includegraphics*[width=0.8\textwidth, keepaspectratio=]{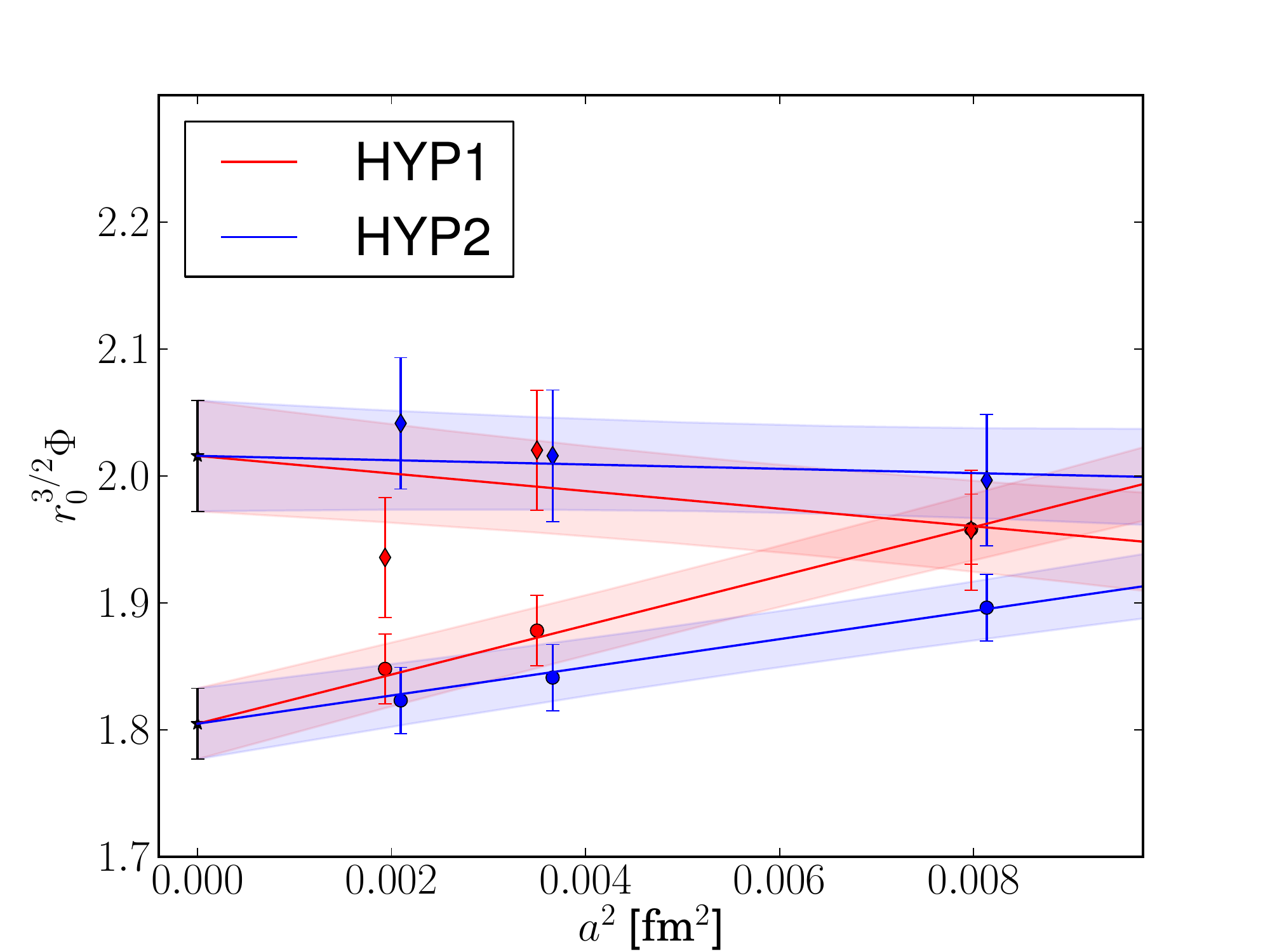}
\end{center}
\caption{\label{figphi1} Extrapolation to the continuum limit of $\Phi_1^{{\rm RGI}}$ (circles) 
and $\Phi_1^{\rm HQET}$ (diamonds).}
\end{figure}
\begin{figure}
\begin{center}
\includegraphics*[width=0.8\textwidth,keepaspectratio=]{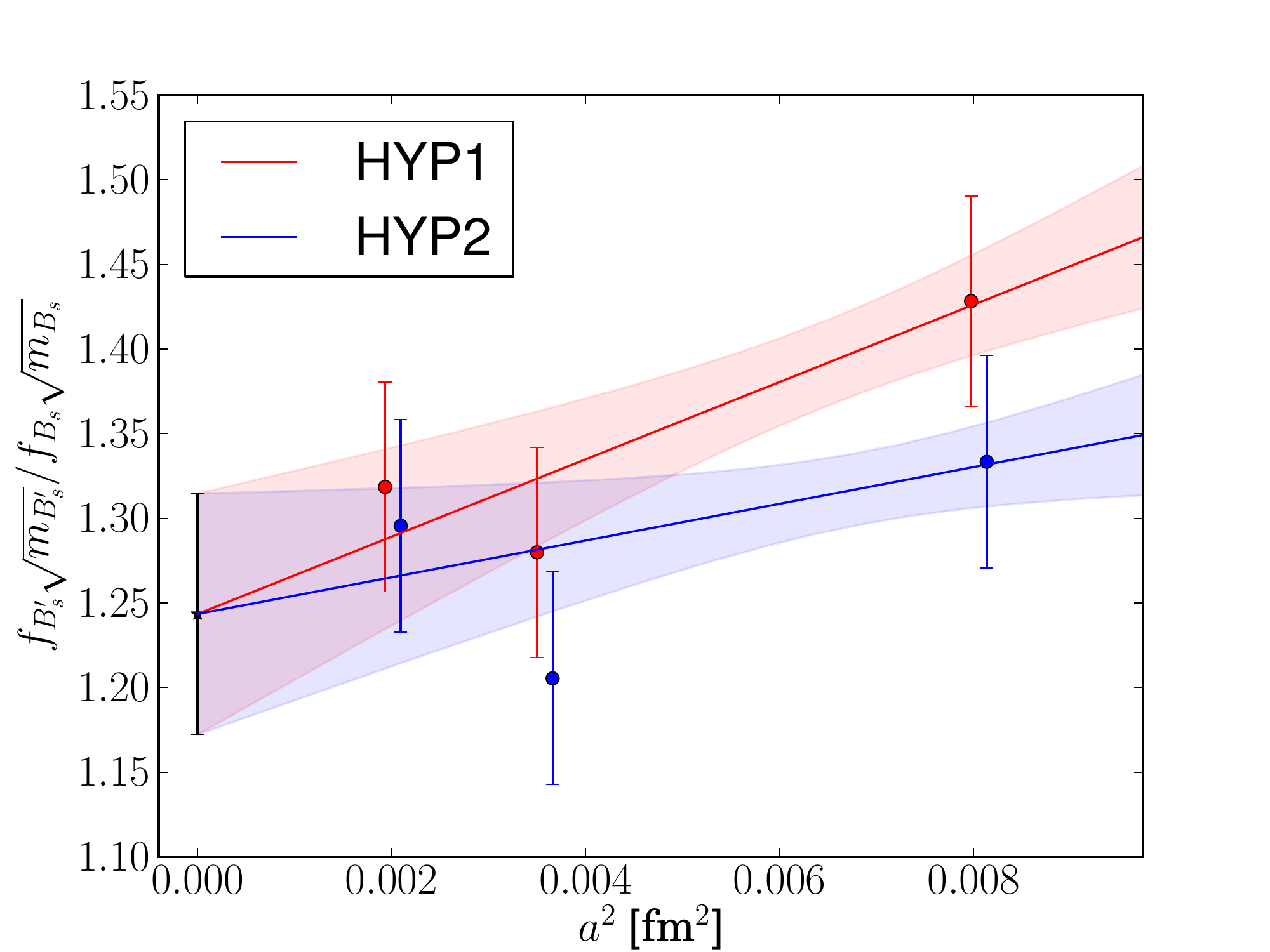}
\end{center}
\caption{\label{figphi2stat} Extrapolation to the continuum limit of 
$f_{\rm B'_{\rm s}}^{\rm stat}\sqrt{m_{\rm B'_{\rm s}}}
/(f_{\rm B_{\rm s}}^{\rm stat}\sqrt{m_{\rm B_{\rm s}}})$.}
\end{figure}
In Figures \ref{figphi1} and \ref{figphi2stat}, we show the continuum
extrapolations of some relevant quantities.
It is easily seen on those plots that the combination of
HYP1 and HYP2 results is legitimate because they point to the
same continuum limit within the errors.

\noindent Numerically we finally get:
\bea\label{fBsstat}
r_0^{3/2}\,f^{\rm stat}_{\rm B_{\rm s}}\sqrt{m_{\rm B_s}} &=&2.14(4)\,,\\
r_0^{3/2}\,f^{\rm HQET}_{\rm B_{\rm s}}\sqrt{m_{\rm B_s}} &=&2.02(5)\,.
\eea
For $r_0=0.5$~fm, our results correspond to
$f^{\rm stat}_{\rm B_{\rm s}}=229(3)$ MeV and
$f^{\rm HQET}_{\rm B_{\rm s}}=216(5)$ MeV,
and for $r_0=0.45$~fm to
$f^{\rm stat}_{\rm B_{\rm s}}=269(4)$ MeV and
$f^{\rm HQET}_{\rm B_{\rm s}}=252(7)$ MeV.

\begin{figure}
\begin{center}
\includegraphics*[width=0.8\textwidth,keepaspectratio=]{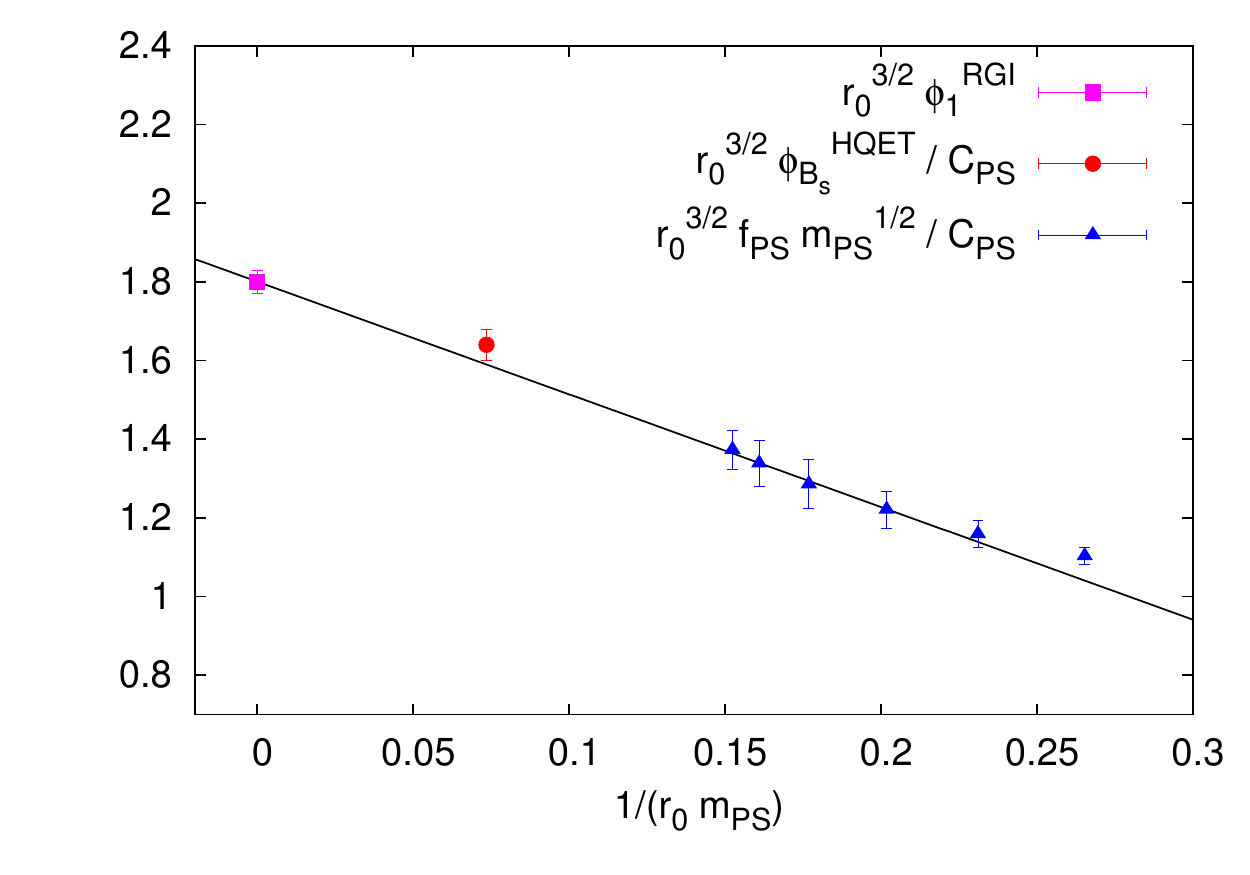}
\end{center}
\caption{\label{figfbinterpol} Comparison between our estimation of
$\Phi^{\rm HQET}_{\rm B_s} = f^{\rm HQET}_{\rm B_s}\,\sqrt{m_{\rm B_s}}$ and the interpolation at $m_{\rm B_s}$ of
$f^{\rm stat}_{\rm B_s}$ and $f_{\rm D_s}$ for a range of $\rm D_s$ masses
\protect\cite{fBsalpha}.
The notations are explained in the text.}
\end{figure}
Although our direct computation of $f_{\rm B_s}$ in HQET avoids
any interpolation (or extrapolation) in the heavy mass,
we show in Figure \ref{figfbinterpol} a comparison of our HQET result with
an interpolation between previous results for $f_{\rm D_s}$
\cite{fBsalpha}
and the static value.
As the decay constant itself does not have a well defined infinite mass
limit, in the figure we plot the quantity
$r_0^{3/2} { {f_{\rm PS}\sqrt{m_{\rm PS}}}
\over { C_{\rm PS}(M/\Lambda_{\overline{\rm MS}}) }} $,
by properly rescaling our non-perturbative result for
$\Phi_1^{\rm HQET}$ (red circle in Figure 4) and the non-perturbative
result in
\cite{fBsalpha}
for the decay constant around the charm quark mass
(blue triangles in Figure 4). The static limit of this
quantity is $r_0^{3/2} \Phi_1^{\rm RGI}$, which we have also
non-perturbatively computed here (purple square in Figure 4).
As explained above we rely on perturbation theory only for the evaluation
of the conversion function $C_{\rm PS} (M/\Lambda_{\overline{\rm MS}})$
relating the RGI matrix elements in static HQET with their counterpart
in QCD defined at a given heavy quark mass
\cite{KurthHeitger2003, HeitgerWennekers2004}. 
Thus, dividing by $C_{\rm PS}(M_{\rm b}/\Lambda_{\overline{\rm MS}})$
compensates for the well-known logarithmic scaling of the
decay constant with the heavy-quark mass
\cite{Shifman:1986sm,Politzer:1988wp}.
One can see that our result is
falling rather well on the straight line expected from 
heavy quark scaling, indicating that the neglected
$\rmO(\minv^n)|_{n \ge 2}$ corrections are small.
We note, however, that this comparison and conclusion rely
on the perturbative evaluation of $C_{\rm PS}$, and that
the associated $\alpha_\mrm{s}(m)^3$ errors are very
difficult to estimate.

Our use of the GEVP method also allows us to extract some
information on the matrix element for $n=2$, i.e. of the first excited
state of the $\rm B_s$ system, for which we obtain
\be
{f^{\rm stat}_{\rm B'_{\rm s}}\sqrt{m_{\rm B'_s}}\over f^{\rm stat}_{\rm B_{\rm s}}\sqrt{m_{\rm B_s}}}= 1.24(7)
\ee
from the ratio ${{p^{\rm stat}_2(1+c_{\rm A}^{\rm stat}p_2^{\rm A^{(1)}})}\over {p^{\rm stat}_1(1+c_{\rm A}^{\rm stat}p_1^{\rm A^{(1)}})} }$
quadratically extrapolated to the continuum limit. The unimproved
version of this quantity (i.e. $p^{\rm stat}_2/p^{\rm stat}_1$), quadratically
extrapolated to the continuum, gives the same result of 1.24(7). 
We have obtained the same qualitative result as
\cite{Burchspectrum2008}
concerning this ratio: it is noticeably larger than 1,
in good qualitative agreement with predictions from  
quark models that become Lorentz covariant in the heavy quark limit
\cite{Morenas1997}
and relativistic quasi-potential quark models
\cite{Ebert:1997nk,Ebert:2002qa},
while other models predict a value less than 1 for this quantity
\cite{Badalian:2007km}.

%% file: s4fBs.tex
\section{Conclusion}\label{secconclusion}

In this paper we have reported on the computation of 
the $\rm B_{\rm s}$ meson decay constant by using
lattice simulations in quenched HQET. Including
$\minv$ corrections introduces power divergences 
$\sim 1/(a\mbeauty)$ which have to be subtracted 
non-perturbatively. These non-perturbative subtractions  
have here been carried out successfully for the first time
in lattice gauge theory computations. 
The necessary couplings of the effective theory had been determined 
non-perturbatively by matching it to QCD 
\cite{hqet:first1}.

Our strategy had already been developed earlier
\cite{strategyfBs}
but its implementation revealed relatively large statistical 
errors in the matrix elements of the 
$\minv$ operators  (not due to the computation of the non-perturbative
parameters of the theory). This shortcoming has now been cured
by exploiting (i) a method based on solving a GEVP
to reduce the systematic errors on bare matrix
elements coming from the contribution of excited states
to correlation functions and (ii) all-to-all propagators to improve the
statistical precision. For example,
at the finest lattice resolution considered,
 we have obtained a result
for the bare static decay constant (HYP2 action), which is three
times more precise than the result in~\cite{fBsalpha} at $\beta=6.45$
where ten times more configurations were analyzed in the Schr\"odinger Functional 
setup.

We used three lattice spacings to extrapolate to the continuum limit. 
With $r_0= 0.5$ fm we have obtained
$f^{\rm stat, N_{\rm f}=0}_{\rm B_{\rm s}} = 229(3)$ MeV and 
$f^{\rm HQET, N_{\rm f}=0}_{\rm B_{\rm s}}= 216(5)$ MeV.
Thus the relative $\minv$ corrections are small as expected from
simple estimates such as $400\,\MeV/\mbeauty$ 
and we have found evidence that $\rmO(\minv^2)$ corrections
are very small. 
In addition, we have shown that the GEVP method
is useful for studying phenomenologically interesting
quantities involving radial excitations of mesonic states,
such as the ratio $f_{\rm B'_{\rm s}}/f_{\rm B_{\rm s}}$.
In this respect we confirm a recent lattice calculation
\cite{Burchspectrum2008}
finding that $f_{\rm B'_{\rm s}}/f_{\rm B_{\rm s}} > 1$ 
at least in the static approximation.

We intend to apply the approach described in this paper to the
computation of $f_{\rm B_{\rm s}}$ and $f_{\rm B}$ on 
dynamical $N_{\rm f}=2$ configurations in the near future.
Note that the problem posed for charm physics on the lattice
by the rapid slowing-down of the topological modes of the gauge fields
with decreasing lattice spacing
\cite{DelDebbio:2002xa,DelDebbio:2004xh,Schaefer:2009xx}
is less relevant in this case, since in HQET we can afford to work 
with coarser lattices.

%% file: fBs.bbl
\providecommand{\href}[2]{#2}\begingroup\raggedright\begin{thebibliography}{10}

\bibitem{Antonelli:2009ws}
M.~Antonelli {\em et~al.}, {\it {Flavor Physics in the Quark Sector}},
  \href{http://xxx.lanl.gov/abs/0907.5386}{{\tt 0907.5386}}.

\bibitem{BRBsmuplusmuminusSM1}
M.~Misiak and J.~Urban, {\it {QCD corrections to FCNC decays mediated by
  Z-penguins and W-boxes}},  {\em Phys. Lett.} {\bf B451} (1999) 161--169,
  [\href{http://xxx.lanl.gov/abs/hep-ph/9901278}{{\tt hep-ph/9901278}}].

\bibitem{BRBsmuplusmuminusSM2}
G.~Buchalla and A.~J. Buras, {\it {The rare decays $K \to \pi \nu \bar{\nu}$,
  $B \to X \nu \bar{\nu}$ and $B \to \ell^+ \ell^-$: An update}},  {\em Nucl.
  Phys.} {\bf B548} (1999) 309--327,
  [\href{http://xxx.lanl.gov/abs/hep-ph/9901288}{{\tt hep-ph/9901288}}].

\bibitem{BRBsmuplusmuminusSM3}
A.~J. Buras, {\it {Relations between $\Delta M(s, d)$ and $B(s, d) \to \mu
  \bar{\mu}$ in models with minimal flavor violation}},  {\em Phys. Lett.} {\bf
  B566} (2003) 115--119, [\href{http://xxx.lanl.gov/abs/hep-ph/0303060}{{\tt
  hep-ph/0303060}}].

\bibitem{BRBsmuplusmuminusD0}
{\bf D0} Collaboration, V.~M. Abazov {\em et~al.}, {\it {Search for the rare
  decay $B_s^0\to \mu^+\mu^-$}},  {\em Phys. Lett.} {\bf B693} (2010) 539--544,
  [\href{http://xxx.lanl.gov/abs/1006.3469}{{\tt 1006.3469}}].

\bibitem{NPBSUSY}
K.~S. Babu and C.~F. Kolda, {\it {Higgs mediated $B^0 \to \mu^{+} \mu^{-}$ in
  minimal supersymmetry}},  {\em Phys. Rev. Lett.} {\bf 84} (2000) 228--231,
  [\href{http://xxx.lanl.gov/abs/hep-ph/9909476}{{\tt hep-ph/9909476}}].

\bibitem{BRLHCbBmuplusmuminus}
{\bf LHCb} Collaboration, P.~Perret {\em et~al.}, {\it {Prospects for New
  Physics in CP Violation and Rare Decays at LHCb}},
  \href{http://xxx.lanl.gov/abs/0901.2856}{{\tt 0901.2856}}.

\bibitem{DellaMorte:2007ny}
M.~Della~Morte, {\it {Standard Model parameters and heavy quarks on the
  lattice}},  {\em PoS} {\bf LAT2007} (2007) 008,
  [\href{http://xxx.lanl.gov/abs/0711.3160}{{\tt 0711.3160}}].

\bibitem{Gamizlattice08}
E.~Gamiz, {\it {Heavy flavour phenomenology from lattice QCD}},  {\em PoS} {\bf
  LATTICE2008} (2008) 014, [\href{http://xxx.lanl.gov/abs/0811.4146}{{\tt
  0811.4146}}].

\bibitem{Eichten:1989zv}
E.~Eichten and B.~R. Hill, {\it {An Effective Field Theory for the Calculation
  of Matrix Elements Involving Heavy Quarks}},  {\em Phys. Lett.} {\bf B234}
  (1990) 511.

\bibitem{Eichten:1990vp}
E.~Eichten and B.~R. Hill, {\it {Static effective field theory: 1/m
  corrections}},  {\em Phys. Lett.} {\bf B243} (1990) 427--431.

\bibitem{hqet:first1}
{\bf ALPHA} Collaboration, B.~Blossier, M.~Della~Morte, N.~Garron, and
  R.~Sommer, {\it {HQET at order $1/m$: I. Non-perturbative parameters in the
  quenched approximation}},  {\em JHEP} {\bf 06} (2010) 002,
  [\href{http://xxx.lanl.gov/abs/1001.4783}{{\tt 1001.4783}}].

\bibitem{strategyfBs}
{\bf ALPHA} Collaboration, B.~Blossier, M.~Della~Morte, N.~Garron, and
  R.~Sommer, {\it {Heavy-light decay constant at the 1/m order of HQET}},  {\em
  PoS} {\bf LAT2007} (2007) 245, [\href{http://xxx.lanl.gov/abs/0710.1553}{{\tt
  0710.1553}}].

\bibitem{alphaGEVP}
{\bf ALPHA} Collaboration, B.~Blossier, M.~Della~Morte, G.~von Hippel,
  T.~Mendes, and R.~Sommer, {\it {On the generalized eigenvalue method for
  energies and matrix elements in lattice field theory}},  {\em JHEP} {\bf 04}
  (2009) 094, [\href{http://xxx.lanl.gov/abs/0902.1265}{{\tt 0902.1265}}].

\bibitem{Kurth:2000ki}
{\bf ALPHA} Collaboration, M.~Kurth and R.~Sommer, {\it {Renormalization and
  O($a$)-improvement of the static axial current}},  {\em Nucl. Phys.} {\bf
  B597} (2001) 488--518, [\href{http://xxx.lanl.gov/abs/hep-lat/0007002}{{\tt
  hep-lat/0007002}}].

\bibitem{gevp:michael}
C.~Michael and I.~Teasdale, {\it Extracting glueball masses from lattice qcd},
  {\em Nucl. Phys.} {\bf B215} (1983) 433--446.

\bibitem{phaseshifts:LW}
M.~{L\"uscher} and U.~Wolff, {\it How to calculate the elastic scattering
  matrix in two-dimensional quantum field theories by numerical simulation},
  {\em Nucl. Phys.} {\bf B339} (1990) 222--252.

\bibitem{Cstatpalombi}
{\bf ALPHA} Collaboration, A.~Grimbach, D.~Guazzini, F.~Knechtli, and
  F.~Palombi, {\it {O(a) improvement of the HYP static axial and vector
  currents at one-loop order of perturbation theory}},  {\em JHEP} {\bf 03}
  (2008) 039, [\href{http://xxx.lanl.gov/abs/0802.0862}{{\tt 0802.0862}}].

\bibitem{HYP}
A.~Hasenfratz and F.~Knechtli, {\it {Flavor symmetry and the static potential
  with hypercubic blocking}},  {\em Phys. Rev.} {\bf D64} (2001) 034504,
  [\href{http://xxx.lanl.gov/abs/hep-lat/0103029}{{\tt hep-lat/0103029}}].

\bibitem{HYP:pot}
A.~Hasenfratz, R.~Hoffmann, and F.~Knechtli, {\it {The static potential with
  hypercubic blocking}},  {\em Nucl. Phys. Proc. Suppl.} {\bf 106} (2002)
  418--420, [\href{http://xxx.lanl.gov/abs/hep-lat/0110168}{{\tt
  hep-lat/0110168}}].

\bibitem{Dellam2005}
{\bf ALPHA} Collaboration, M.~Della~Morte, A.~Shindler, and R.~Sommer, {\it {On
  lattice actions for static quarks}},  {\em JHEP} {\bf 08} (2005) 051,
  [\href{http://xxx.lanl.gov/abs/hep-lat/0506008}{{\tt hep-lat/0506008}}].

\bibitem{impr:SW}
B.~Sheikholeslami and R.~Wohlert, {\it Improved continuum limit lattice action
  for {QCD} with {W}ilson fermions},  {\em Nucl. Phys.} {\bf B259} (1985) 572.

\bibitem{impr:pap3}
M.~L\"uscher, S.~Sint, R.~Sommer, P.~Weisz, and U.~Wolff, {\it
  {Non-perturbative O(a) improvement of lattice QCD}},  {\em Nucl. Phys.} {\bf
  B491} (1997) 323--343, [\href{http://xxx.lanl.gov/abs/hep-lat/9609035}{{\tt
  hep-lat/9609035}}].

\bibitem{dublin}
{\bf TrinLat} Collaboration, J.~Foley {\em et~al.}, {\it {Practical all-to-all
  propagators for lattice QCD}},  {\em Comput. Phys. Commun.} {\bf 172} (2005)
  145--162, [\href{http://xxx.lanl.gov/abs/hep-lat/0505023}{{\tt
  hep-lat/0505023}}].

\bibitem{hqet:first3}
{\bf ALPHA} Collaboration, B.~Blossier {\em et~al.}, {\it {HQET at order $1/m$:
  II. Spectroscopy in the quenched approximation}},  {\em JHEP} {\bf 05} (2010)
  074, [\href{http://xxx.lanl.gov/abs/1004.2661}{{\tt 1004.2661}}].

\bibitem{r0formula}
{\bf ALPHA} Collaboration, M.~Guagnelli, R.~Sommer, and H.~Wittig, {\it
  {Precision computation of a low-energy reference scale in quenched lattice
  QCD}},  {\em Nucl. Phys.} {\bf B535} (1998) 389--402,
  [\href{http://xxx.lanl.gov/abs/hep-lat/9806005}{{\tt hep-lat/9806005}}].

\bibitem{mbar:pap3}
{\bf ALPHA} Collaboration, J.~Garden, J.~Heitger, R.~Sommer, and H.~Wittig,
  {\it Precision computation of the strange quark's mass in quenched {QCD}},
  {\em Nucl. Phys.} {\bf B571} (2000) 237--256,
  [\href{http://xxx.lanl.gov/abs/hep-lat/9906013}{{\tt hep-lat/9906013}}].

\bibitem{wavef:wupp1}
S.~{G\"usken} {\em et~al.}, {\it {Nonsinglet Axial Vector Couplings of the
  Baryon Octet in Lattice QCD}},  {\em Phys. Lett.} {\bf B227} (1989) 266.

\bibitem{smear:ape}
{\bf APE} Collaboration, M.~Albanese {\em et~al.}, {\it Glueball masses and
  string tension in lattice {QCD}},  {\em Phys. Lett.} {\bf 192B} (1987) 163.

\bibitem{Basak:2005gi}
S.~Basak {\em et~al.}, {\it {Combining Quark and Link Smearing to Improve
  Extended Baryon Operators}},  {\em PoS} {\bf LAT2005} (2006) 076,
  [\href{http://xxx.lanl.gov/abs/hep-lat/0509179}{{\tt hep-lat/0509179}}].

\bibitem{KurthHeitger2003}
{\bf ALPHA} Collaboration, J.~Heitger, M.~Kurth, and R.~Sommer, {\it
  {Non-perturbative renormalization of the static axial current in quenched
  QCD}},  {\em Nucl. Phys.} {\bf B669} (2003) 173--206,
  [\href{http://xxx.lanl.gov/abs/hep-lat/0302019}{{\tt hep-lat/0302019}}].

\bibitem{HeitgerWennekers2004}
{\bf ALPHA} Collaboration, J.~Heitger, A.~J\"uttner, R.~Sommer, and
  J.~Wennekers, {\it {Non-perturbative tests of heavy quark effective theory}},
   {\em JHEP} {\bf 11} (2004) 048,
  [\href{http://xxx.lanl.gov/abs/hep-ph/0407227}{{\tt hep-ph/0407227}}].

\bibitem{Chetyrkin:2003vi}
K.~G. Chetyrkin and A.~G. Grozin, {\it {Three-loop anomalous dimension of the
  heavy-light quark current in HQET}},  {\em Nucl. Phys.} {\bf B666} (2003)
  289--302, [\href{http://xxx.lanl.gov/abs/hep-ph/0303113}{{\tt
  hep-ph/0303113}}].

\bibitem{Bekavac:2009zc}
S.~Bekavac {\em et~al.}, {\it {Matching QCD and HQET heavy-light currents at
  three loops}},  {\em Nucl. Phys.} {\bf B833} (2010) 46--63,
  [\href{http://xxx.lanl.gov/abs/0911.3356}{{\tt 0911.3356}}].

\bibitem{fBsalpha}
M.~Della~Morte {\em et~al.}, {\it {Heavy-strange meson decay constants in the
  continuum limit of quenched QCD}},  {\em JHEP} {\bf 02} (2008) 078,
  [\href{http://xxx.lanl.gov/abs/0710.2201}{{\tt 0710.2201}}].

\bibitem{Shifman:1986sm}
M.~A. Shifman and M.~B. Voloshin, {\it {On Annihilation of Mesons Built from
  Heavy and Light Quark and $\overline{B^0} - B^0$ Oscillations}},  {\em Sov.
  J. Nucl. Phys.} {\bf 45} (1987) 292.

\bibitem{Politzer:1988wp}
H.~D. Politzer and M.~B. Wise, {\it {Leading Logarithms of Heavy Quark Masses
  in Processes with Light and Heavy Quarks}},  {\em Phys. Lett.} {\bf B206}
  (1988) 681.

\bibitem{Burchspectrum2008}
T.~Burch, C.~Hagen, C.~B. Lang, M.~Limmer, and A.~Sch\"afer, {\it {Excitations
  of single-beauty hadrons}},  {\em Phys. Rev.} {\bf D79} (2009) 014504,
  [\href{http://xxx.lanl.gov/abs/0809.1103}{{\tt 0809.1103}}].

\bibitem{Morenas1997}
V.~Mor\'enas, A.~Le~Yaouanc, L.~Oliver, O.~P\`ene, and J.~C. Raynal, {\it
  {Decay constants in the heavy quark limit in models \`a la Bakamjian and
  Thomas}},  {\em Phys. Rev.} {\bf D58} (1998) 114019,
  [\href{http://xxx.lanl.gov/abs/hep-ph/9710298}{{\tt hep-ph/9710298}}].

\bibitem{Ebert:1997nk}
D.~Ebert, V.~O. Galkin, and R.~N. Faustov, {\it {Mass spectrum of orbitally and
  radially excited heavy- light mesons in the relativistic quark model}},  {\em
  Phys. Rev.} {\bf D57} (1998) 5663--5669,
  [\href{http://xxx.lanl.gov/abs/hep-ph/9712318}{{\tt hep-ph/9712318}}].

\bibitem{Ebert:2002qa}
D.~Ebert, R.~N. Faustov, and V.~O. Galkin, {\it {Decay constants of heavy-light
  mesons in the relativistic quark model}},  {\em Mod. Phys. Lett.} {\bf A17}
  (2002) 803--808, [\href{http://xxx.lanl.gov/abs/hep-ph/0204167}{{\tt
  hep-ph/0204167}}].

\bibitem{Badalian:2007km}
A.~M. Badalian, B.~L.~G. Bakker, and Y.~A. Simonov, {\it {Decay constants of
  the heavy-light mesons from the field correlator method}},  {\em Phys. Rev.}
  {\bf D75} (2007) 116001, [\href{http://xxx.lanl.gov/abs/hep-ph/0702157}{{\tt
  hep-ph/0702157}}].

\bibitem{DelDebbio:2002xa}
L.~Del~Debbio, H.~Panagopoulos, and E.~Vicari, {\it {theta dependence of
  SU($N$) gauge theories}},  {\em JHEP} {\bf 0208} (2002) 044,
  [\href{http://xxx.lanl.gov/abs/hep-th/0204125}{{\tt hep-th/0204125}}].

\bibitem{DelDebbio:2004xh}
L.~Del~Debbio, G.~M. Manca, and E.~Vicari, {\it {Critical slowing down of
  topological modes}},  {\em Phys. Lett.} {\bf B594} (2004) 315--323,
  [\href{http://xxx.lanl.gov/abs/hep-lat/0403001}{{\tt hep-lat/0403001}}].

\bibitem{Schaefer:2009xx}
S.~Schaefer, R.~Sommer, and F.~Virotta, {\it {Investigating the critical
  slowing down of QCD simulations}},
  \href{http://xxx.lanl.gov/abs/0910.1465}{{\tt 0910.1465}}.

\end{thebibliography}\endgroup
